\input harvmac
\input epsf
\def \sign{ \mathop{ \rm sign}\nolimits}
\def \trace{ \mathop{ \rm trace}\nolimits}
\Title{\vbox{\hbox to 5cm{ \hfill  DFTUZ /95/23}
                   \hbox to 5cm{ \hfill  hep-th/9603046}
         }}
{\vbox{\centerline{A method for solve integrable $A_2$ spin chains }
\vskip2pt\centerline{combining different representations }}}
\centerline{J. Abad and M. Rios}
\centerline{Departamento de F\'{\i}sica Te\'{o}rica, Facultad de Ciencias,}
\centerline{Universidad de Zaragoza, 50009 Zaragoza, Spain}
\bigskip
\bigskip
\vskip .3in
\centerline{ \tenbf Abstract}
A non homogeneous spin chain in the representations $ \{3 \}$ and $ \{3^*\}$
of $A_2$ is analyzed. We find that the naive nested Bethe ansatz is not
applicable to this case. A method inspired in the nested Bethe ansatz, that can
be applied to more general cases, is developed for that chain. The solution for
the eigenvalues of the trace of the monodromy matrix is given as two coupled
Bethe equations different from that for a homogeneous chain. A conjecture about
the form of the solutions for more general chains is presented
\bigskip
\bigskip
\Date{}

\vfill
\eject

Integrable magnetic chains are interesting physical systems with a rich
mathematical structure. Homogeneous integrable spin chains with spin $1/2$ and
higher have been found and solved \ref\rri{H. M. Babujian, Phys. Lett. A 90
(1982) 479.}%
\nref\rrii{L.A. Takhtajan, Phys.Lett. A 87 (1982) 479.}%
\nref\rriii{A. N. Kirillov and N. Y. Reshetikhin, J. Phys. A 20 (1987) 1565.}%
\nref\rriv{H. M. Babujian and A. M. Tsvelick, Nucl. Phys. B 265 (1986) 24.}%
--\ref\rrv{A. B. Zamolodchikov and V. A. Fateev, Sov. J. Nucl. Phys. 32 (1980)
298.}.
Other interesting systems are non-homogenous chains combining  two different
kinds of spin states in the sites. De Vega and Woynarovich  \ref\ri{ H.J. de
Vega and  F. Woynarovich, J.  Phys. A 25 (1992) 4499.} pioneered  the work in
these systems, whose thermodynamic limit  was obtained by de Vega, L.
Mezincescu and R.I. Nepomechie \ref\rii{ H.J. de Vega, L. Mezincescu and R.I.
Nepomechie, Phys. Rev. B 49 (1994) 13223.} (see too
\ref\rrvi{S. R. Aladin and M. J. Martins, J. Phys. A 26 (1993), l529 and J.
Phys. A 26 (1993) 7287.}%
\nref\rrvii{ M. J. Martins, J. Phys. A 26 (1993) 7301.}%
-\ref\rrviii{ H.J. de Vega, L. Mezincescu and R.I. Nepomechie, J. Mod. Phys. B
8 (1994) 3473})%
. These authors considered an alternating chain with spin 1/2 and spin 1
associated to a $su(2)$ algebra. Following the quantum inverse scattering
method (QISM) and using the Yang Baxter equation (YBE) \ref\riii{L. D. Fadeev,
Sov. Sc. Rev. Math Phys. C1 (1981) 107  \semi
V.E. Korepin, N.M. Bogoliubov and A.G. Izergin, Quantum Inverse Scattering
Method and Correlation Functions. Cambridge University Press (1991).}, they
found integrable hamiltonians that present terms coupling pairs of neighboring
sites and others coupling three neighboring spin sites. The method introduces a
monodromy matrix on an auxiliary space whose elements are tensors on the state
space of the chain, tensorial product of the site spaces. The spectrum and
eigenstates of the trace of the monodromy matrix are obtained by using the
Bethe ansatz (BA) method and solving the set of equations, known as Bethe
equations, derived by this method.

In a homogeneous system with the associated algebra of higher rank
(\ref\rrx{M. Jimbo, Commun. Math. Phys. 102 (1986) 537.}%
\nref\rrxi{V. V. Bazhanov, Phys. Lett B 159 (1985), 321 and Commun. Math. Phys.
  113 (1987) 471.}%
\nref\rrxii{M. Jimbo, T. Miwa and M. Okado, Mod. Phys. Lett. B1 (1987) 73.}%
-\ref\rrxiii{H. J.  de Vega and E. Lopez, Phys. Rev. Lett. 67 (1981) 489; Nucl.
Phys. B 362 (1991) 261; Preprint LPTHE 91/29.})%
, the solutions are obtained by applying the BA method successively
$(d-1)$-times, $d$ being the dimension of the spin. This method is know as
nested Bethe ansatz (NBA) \ref\riv{H.J. de Vega, J. Mod. Phys. A4 (1989) 2371
\semi
H.J. de Vega, Nucl. Phys. B (Proc. nad Suppl.) 18A (1990) 229\semi
J. Abad and M. Rios, Univ. of Zaragoza preprint DFTUZ 94-11 (1994).}.

In this paper, we are going to solve a non-homogeneous chain with spin
associated to the algebra $su(3)$ and states of the sites alternating between
the elementary representations $\{3\}$ and  $\{3^*\}$. In order to obtain
hamiltonians associated to alternating chains, we have made an extension of the
method used in ref. \ri\ for systems where $P$ and $T$ symmetry are not
conserved . The diagonalization of our hamiltonians is not possible by the
usual NBA method. The reason why is that we have not a reference state, which
would be taken as pseudovacuum annihilated by the elements under the diagonal
of  the monodromy matrix, to build the eigenstates of the diagonal elements of
the monodromy matrix  whose trace is related to the hamiltonian.

In the homogeneous case, the eigenstates are built by the action of the
elements above the diagonal of the monodromy matrix on the pseudovacuum state;
this fact makes the charges of $su(3)$ be related to the number of applied
operators, which can be considered as a particle number. In the non-homogeneous
case this does not happen anymore, since there are site states with charges of
different sign which belong to different representations and therefore there
are states of the chain with the same total charge and different number of
particles. This makes the usual NBA method non applicable.

The method we are going to present is inspired in NBA one and in the method
given for solutions of $O(2N)$ symmetric theories in ref. %
\ref\rrix{H.J. de Vega and M. Karowski, Nucl Phys. B 280 (1987) 225.}%
, but it has important differences with them. In a first step, we look for a
subspace of states of the chain that are eigenstates of the element $T_{1,1}$
of the monodromy matrix and are annihilate by the elements $T_{j, 1}$ $(j >1)$,
the whole subspace being invariant under application of  $T_{i, j}$ $(i, j
>1)$. On this subspace, we build the eigenstates of the trace of $T$ by the
application of operators $T_{1, j}$ $(j >1)$ and obtain the eigenvalues in
terms of the restriction of  $T $ to one less dimension. The procedure should
be repeated to end with a monodromy matrix of dimension one. As we see, our
method can be considered as an extension of the NBA one.

In this paper we start by showing the general features of a non-homogeneous
chain \ri. Then develop our method in an alternating chain based in the
representations
$\{3\}$ and  $\{3^*\}$ of $su(3)$, but the method has a general application
that we will applied in a forthcoming paper.

The YBE is usually written
\eqn\ei{  R(\theta-\theta') \cdot [t(\theta)\otimes  t(\theta')]=
 [t(\theta')\otimes  t(\theta)] \cdot R(\theta-\theta'),}
with
\eqn\eii{  [t_{a,b} (\theta)]_{c,d} =R_{c,a}^{b,d} (\theta)}
and the product $\otimes$ being understood, as usual, in the indices $c$ and
$d$ which lie in the site space and the $\cdot$ product in the space tensorial
product of the two auxiliary spaces in which lie the indices $a$ and $b$ of $t$
and all indices of $R$. Both operators are represented in fig. 1(a).
\bigskip
\centerline{\epsfxsize=10cm  \epsfbox{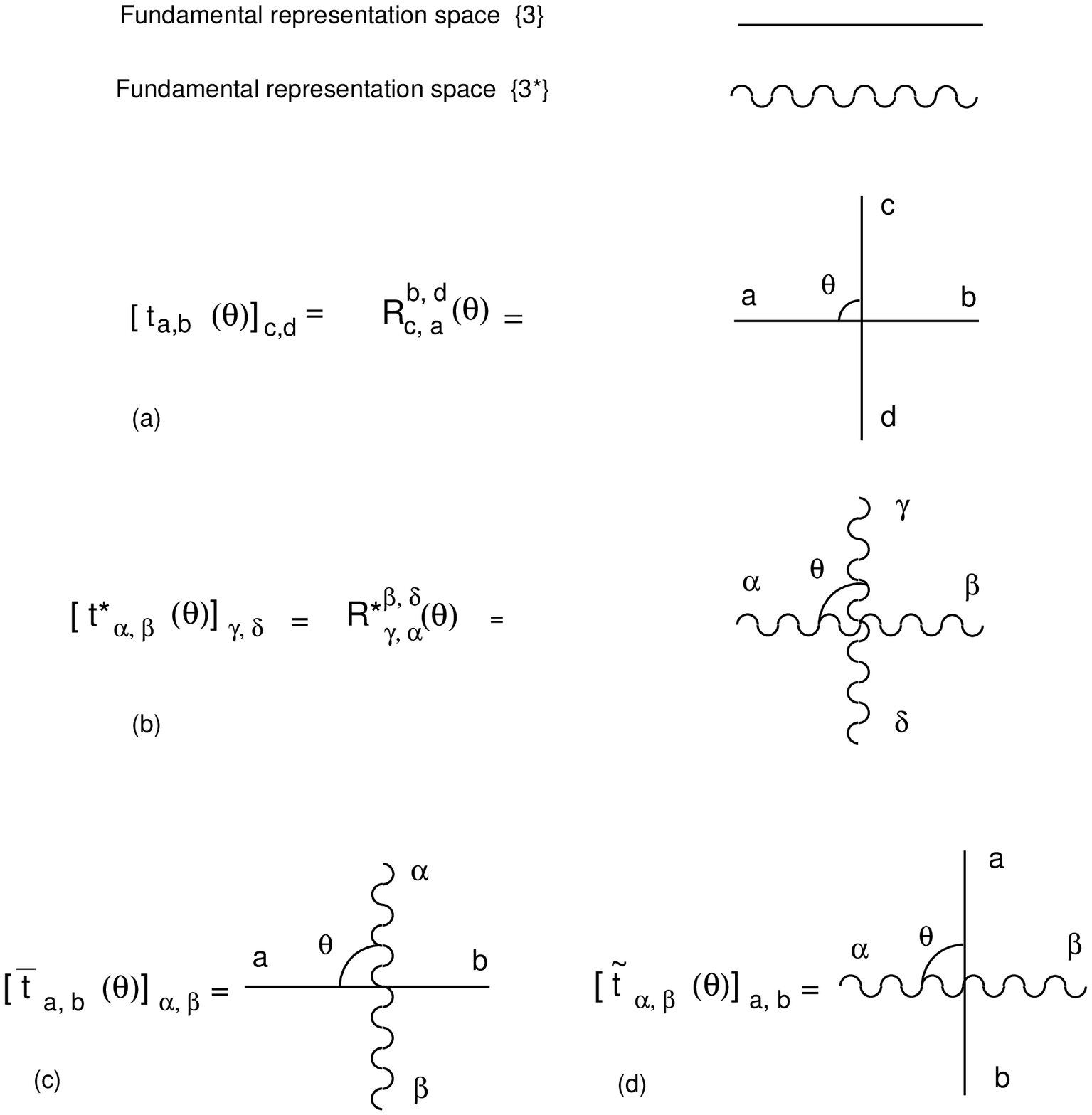}}
\centerline{Fig. 1}
\bigskip

There is another solution to \ei\ with the same  $R$-matrix and a $t$-matrix,
with the site space in  other representation, that we call $\bar{t}$ and is
represented in fig. 1(c) with self-explanatory notation. The YBE \ei\ for these
operators is represented in fig. 2.
\bigskip
\centerline{\epsfxsize=10cm  \epsfbox{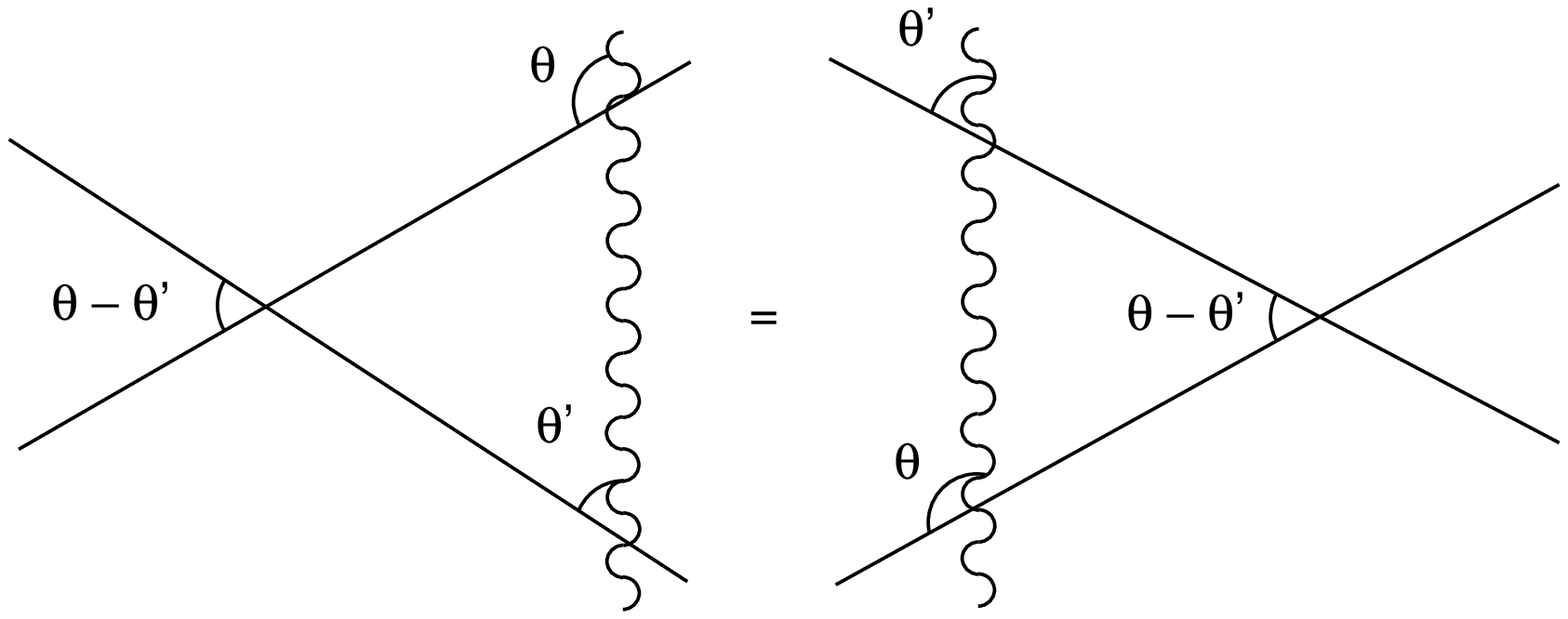}}
\centerline{Fig.2 }
\bigskip

As said before, the key of the method is to build the monodromy operator as a
product of $t$ and $\bar{t}$ matrices, one per site, in the indices of the
auxiliary space. The method could be applied to any distribution of the spin
representations in the chain. Here, we are considering an alternating chain.
So, we have for the monodromy operator $T^{(alt)}$
\eqn\eiii{
T^{(alt)}_{a,b} (\theta,\alpha)= t^{(1)}_{a,a_1}(\theta)\bar{t}
^{(2)}_{a_1,a_2}
(\theta+\alpha) \ldots
t^{(2N-1)}_{a_{2N-2},a_{2N-1}}(\theta)\bar{t} ^{(2N)}_{a_{2N-1,b}}
(\theta+\alpha),
}
is graphically expressed in fig. 3(a)
\bigskip
\centerline{\epsfxsize=10cm  \epsfbox{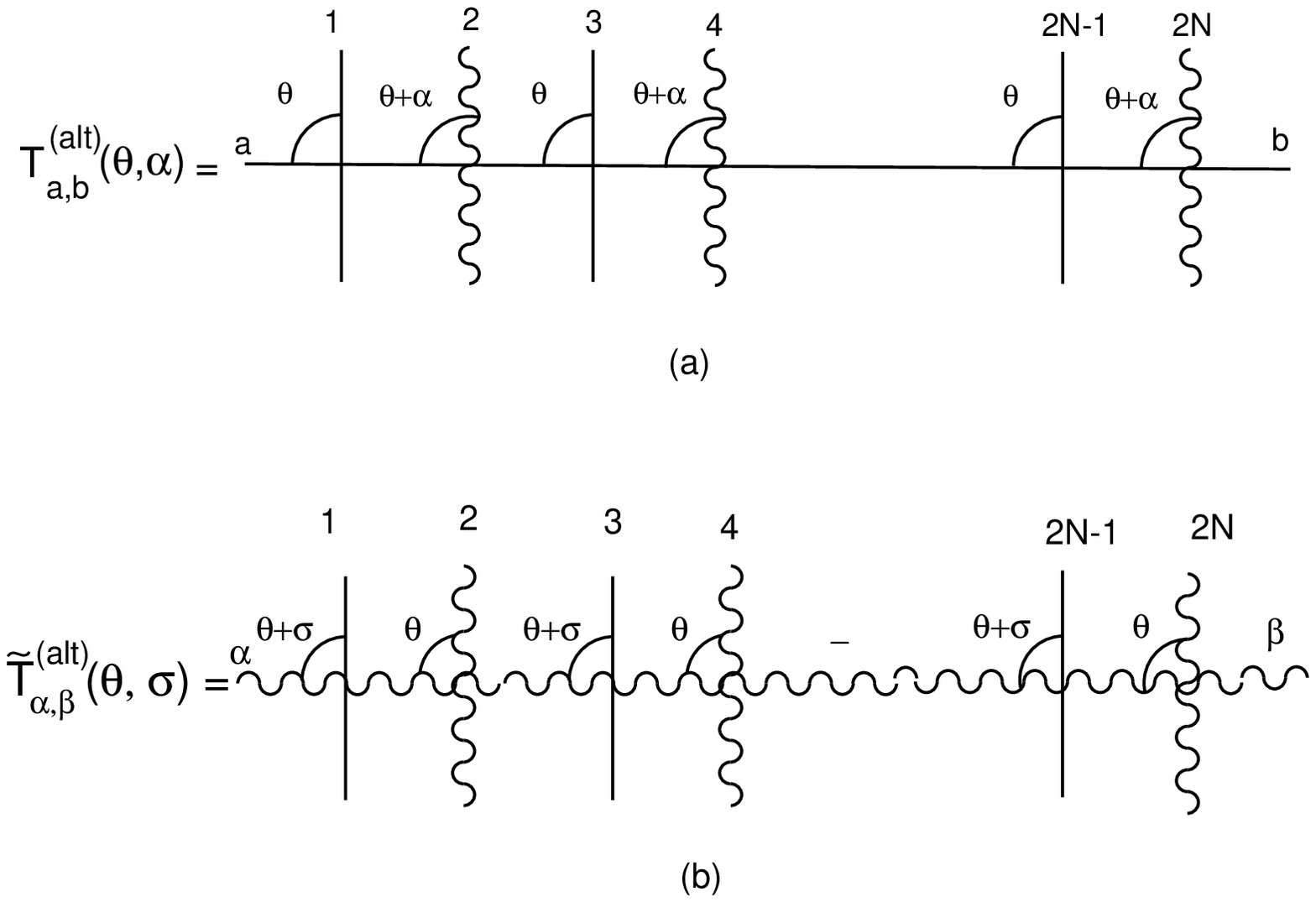}}
\centerline{Fig. 3}
\bigskip
\noindent This operator, due to \ei, fulfills the YBE.

In the same way we have obtained this system, we can considered an analogous
system by  interchanging the representations $\{3\}$ and $\{3^*\}$. The
corresponding $R^*$, $t^*$ and $\tilde{t}$ operators, represented in fig. 1(b)
and (d), satisfy
\eqn\eiv{  R^*(\theta-\theta') \cdot [t^*(\theta)\otimes  t^*(\theta')]=
 [t^*(\theta')\otimes  t^*(\theta)] \cdot R^*(\theta-\theta'),}
with
\eqn\ev{  [t_{a,b} ^*(\theta)]_{c,d} =R_{c,a}^{* b,d} (\theta).}
The corresponding monodromy matrix,  $\tilde{T}^{(alt)}$, is represented in
fig. 3(b).

In the most general case, the $R$ -matrix has the following properties

i) It is $PT$-symmetric
\eqna\eav
$$\eqalignno{R^{c,d}_{a,b}(\theta)&=R^{a,b}_{c,d}(\theta),&\eav a \cr
R^{*}{}^{\gamma,\delta}_{ \alpha,\beta}(\theta)&=R^{ *}{}^{ \alpha,\beta}_{
\gamma,\delta}
(\theta).&\eav b \cr}
$$

ii) It is unitary
\eqna\eavi
$$\eqalignno{&R^{c,d}_{a,b}(\theta) R^{e,f}_{c,d}(-\theta) =
\rho(\theta) \delta_{a,e} \delta_{b,f}, &\eavi a \cr
&R^{*}{}^{\gamma,\delta}_{ \alpha,\beta}(\theta) R^{*}{}^{\mu,\nu}_{
\gamma,\delta}(-\theta)=\rho^*(\theta) \delta_{\alpha,\mu}
\delta_{\beta,\upsilon}. &\eavi b \cr}
$$

iii) The values for the spectral parameter $\theta=0$ are $R(0)= c_0 I$ and
$R^*(0)= c^*_0 I  $

iv) A matrix $M$ and a constant $\eta$ exist such that
\eqn\eaviii{
R^{c,d}_{a,b}(\theta) M_{b,e} R^{g,e}_{f,d}(-\theta-2\eta) M^{-1}_{f,h}\propto
\delta_{a,g}\delta_{c,h}.
}

v) The $t$-matrices verify
\eqn\eaix{
[\bar{t}_{a,b}(\theta)]_{\alpha,\beta}  [
\tilde{t}_{\beta,\gamma}(-\theta)]_{b,c}=\tilde{\rho} (\theta)
\delta_{a,c}\delta_{\alpha,\gamma}.
}

The hamiltonian is obtained by taking the trace of  $T^{(alt)}$ in the
auxiliary space
\eqn\evi{\tau^{(alt)}(\theta,\alpha) = T^{(alt)}_{a,a}(\theta,\alpha),}
and doing
\eqn\evii{
H={d \over d\theta}{\ln{\tau^{(alt)}(\theta,\alpha)}\bigl|_{\theta=0}}.
}
The successive derivatives of $\tau^{(alt)}$ give a complete set of commuting
operators that make the system integrable.

Unlike in the homogeneous case, the hamiltonian has pieces that involve two and
three neighboring sites
\eqn\eviii{
H={1 \over{\tilde{\rho}(\alpha)}}
\sum_{i=1 \atop i odd}^{2N-1}{h^{(1)}_{i, i+1}} +
{1 \over{c_0 \tilde{\rho}(\alpha)}}
\sum_{i=1 \atop i odd}^{2N-2}{h^{(2)}_{i, i+1,i+2}},
}
with
\eqn\eix{
(h^{(1)}_{i, i+1})_{a, \beta; b, \gamma}=[ \dot{\bar{t}}_{a, c}
(\alpha)]_{\beta, \delta} [\tilde{t}_{\delta, \gamma}(-\alpha)]_{c, b}
}
and
\eqn\ex{
(h^{(2)}_{i, i+1, i+2})_{a, \beta, c;   b, \gamma, d}=[\bar{t}_{a,
e}(\alpha)]_{\beta, \delta} [\dot{t}_{e, d}(0)]_{c, f} [\tilde{t}_{\delta,
\gamma}(-\alpha)]_{f, b}
}
where  $\dot{t}$ means $d t/d\theta$.
%
%
%

In the same way, we can obtain another system by interchanging the
representations $\{3\}$ and $\{3^*\}$. The hamiltonian $\tilde{H}$ so obtained
commutes with $H$ and both of them can be simultaneously diagonalized.

The chain we are considering, the simplest one in a higher rank algebra,
combines alternating the two elementary representations of $su(3)$. In fig. 1,
we associate the solid line to the $\{3\}$ representation and the wavy line to
the $\{3^*\}$. The $t$ matrix is \riv
\eqn\exi{
\eqalign{
t(\theta) =& \sinh{( {3 \over 2} \theta +\gamma) }  \sum_{i=1}^{3}{
e_{i,i}\otimes e_{i,i} ^s}+
\sinh{({3 \over 2}  \theta)} \sum_{i,j=1 \atop i \neq j}^{3}{ e_{i,i}\otimes
e_{j,j}^s }    \cr &+
\sinh{(\gamma)}  \sum_{i,j=1 \atop i \neq j}^{3} {    exp{[(i-j-{n \over 2}
\sign {(i-j)}) \theta]}
 e_{i,j}\otimes e_{j,i}^s   },\cr}
}
where  the superindex $s$ means site space and, in the $\{3\}$ representation,
the $e_{i,j}$ are the matrices $(e_{i,j})_{l,m}=\delta_{i,l} \delta_{j,m}$ .

The $t$ operator can be written in matrix form
\eqn\exii{
t(\theta)=
\left (\matrix{
a & 0 & 0 & 0 & 0 & 0 & 0 & 0 & 0 \cr
0 & b & 0 & c & 0 & 0 & 0 & 0 & 0 \cr
0 & 0 & b & 0 & 0 & 0 & d & 0 & 0 \cr
0 & d & 0 & b & 0 & 0 & 0 & 0 & 0 \cr
0 & 0 & 0 & 0 & a & 0 & 0 & 0 & 0 \cr
0 & 0 & 0 & 0 & 0 & b & 0 & c & 0 \cr
0 & 0 & c & 0 & 0 & 0 & b & 0 & 0 \cr
0 & 0 & 0 & 0 & 0 & d & 0 & b & 0 \cr
0 & 0 & 0 & 0 & 0 & 0 & 0 & 0 & a \cr
}\right ),
}
with
\eqna\exiii{
$$\eqalignno{
&a(\theta)=\sinh{({3 \over 2} \theta+\gamma)},&\exiii a\cr
&b(\theta)= \sinh{({3 \over 2} \theta)} ,&\exiii b \cr
&c(\theta)=\sinh{(\gamma)}  e^{{ \theta\over 2}},&\exiii c \cr
&d(\theta)=\sinh{(\gamma)} e^{{ -\theta\over 2}}.&\exiii d
\cr
}
$$}
In the same way, we can write
\eqn\exiv{
\bar{t}(\theta)=
\left (\matrix{
\bar{a }& 0 & 0 & \bar{c}& 0 & 0 & 0 & 0 &  \bar{d } \cr
0 &\bar{ b} & 0 & 0 & 0 & 0 & 0 & 0 & 0 \cr
0 & 0 & \bar{ b} & 0 & 0 & 0 & 0 & 0 & 0 \cr
0 & 0 & 0 &\bar{ b} & 0 & 0 & 0 & 0 & 0 \cr
\bar{d } & 0 & 0 & 0 & \bar{a} & 0 & 0 & 0 & \bar{c} \cr
0 & 0 & 0 & 0 & 0 & \bar{b }& 0 & 0 & 0 \cr
0 & 0 & 0 & 0 & 0 & 0 & \bar{b} & 0 & 0 \cr
0 & 0 & 0 & 0 & 0 & 0 & 0 &\bar{ b} & 0 \cr
\bar{c} & 0 & 0 & 0 &  \bar{d } & 0 & 0 & 0 &\bar{ a }\cr
}\right ),
}
with
\eqna\exv{
$$\eqalignno{
&\bar{a}(\theta)=\sinh{({3 \over 2} \theta+{\gamma \over 2})},&\exv a\cr
&\bar{b}(\theta)= \sinh{({3 \over 2} (\theta+\gamma))} ,&\exv b \cr
&\bar{c}(\theta)=-\sinh{(\gamma)}  e^{{ (\theta+\gamma)\over 2}},&\exv c \cr
&\bar{d}(\theta)=-\sinh{(\gamma)} e^{{ -(\theta+\gamma)\over 2}}.&\exv d
\cr
}
$$}

The monodromy matrix $T^{alt}$ can be written  as a matrix in the auxiliary
space,
\eqn\exvi{
T^{alt}(\theta)=
\pmatrix{
A(\theta) & B_2 (\theta) & B_3 (\theta) \cr
C_2 (\theta)& D_{2,2}(\theta) & D_{2,3}(\theta) \cr
C_3 (\theta)& D_{3,2}(\theta) & D_{3,3}(\theta) \cr
},
}
whose elements are operators in the tensorial product of the site spaces.

The YBE for  $T^{alt}$, can be written in terms of its components
\eqna\exvii{
$$\eqalignno{
&B(\theta) \otimes B(\theta')= {R}^{(2)} (\theta-\theta') \cdot\bigl(
B(\theta')\otimes B(\theta) \bigr) = \bigl( B(\theta')\otimes B(\theta)
\bigr)\cdot  {R}^{(2)} (\theta-\theta'), & \exvii a \cr
&A(\theta) B(\theta')=g(\theta'-\theta) B(\theta')  A(\theta) -B(\theta)
A(\theta') \cdot {\tilde{r}}^{(2)}(\theta'-\theta),& \exvii b \cr
&D(\theta) \otimes B(\theta')=g(\theta-\theta') (B(v) \otimes D(\theta))
\cdot{R}^{(2)} (\theta-\theta') - B(\theta) \otimes  ({r}^{(2)}(\theta-\theta')
\cdot D(\theta') ), &\cr & &\exvii c \cr
}
$$}
where
\eqn\exviii{
{R}^{(2)}(\theta) =
\left (\matrix{
1 & 0 & 0 & 0 \cr
0 & d & b & 0 \cr
0 & b & c & 0 \cr
0 & 0 & 0 & 1 \cr
}\right ) ,\qquad
{r}^{(2)}(\theta)=\pmatrix{
{h}_{-}& 0 \cr
0 & {h}_{+}\cr
},\qquad
{\tilde{r}}^{(2)}(\theta)=
\pmatrix{
{h}_{+} & 0 \cr
0 & {h}_{-} \cr
},
}
and
\eqn\exviiir{
g(\theta)={ 1 \over b(\theta) },\qquad {h}_{+}(\theta)= {c(\theta) \over
b(\theta)}, \qquad {h}_{-}(\theta)= {d(\theta) \over b(\theta)}.
}

For the site states, we use the notation
\eqn\exix{
u=\left (\matrix{
1\cr
0\cr
0\cr
}\right ),\quad
d=\left (\matrix{
0\cr
1\cr
0\cr
}\right ),\quad
s=\left (\matrix{
0\cr
0\cr
1\cr
}\right ),\quad
\bar{u}=\left (\matrix{
0\cr
0\cr
1\cr
}\right ),\quad
\bar{d}=\left (\matrix{
0\cr
1\cr
0\cr
}\right ),\quad
\bar{s}=\left (\matrix{
1\cr
0\cr
0\cr
}\right ),
}

Inspired in the NBA method, we look for an eigenstate of $A$ that serves as a
pseudovacuum. Firstly we group two neighbor sites and call
$w_i$ the subspace generated by the vectors $\mid u, \bar{s} >$ and $\mid u,
\bar{d} >$ in the two site space formed with the $2i$-th and $(2i+1)$-th sites.
Then, we build the subspace
\eqn\exx{
\Omega = w_1 \otimes w_2\otimes \cdots  \otimes w_N
}
in the total space of states of a chain with $2N$ sites.

In a nonhomogeneous chain, we have not a state $\parallel s>$ such that $D_{i,
j}\parallel s>\propto\delta_{i,j} \parallel s>$. For this reason, the NBA
method can not be used. Our method, instead,  starts with a state $\parallel 1>
\in  \Omega$ verifying
\eqna\exxi{
$$\eqalignno{
&A(\theta) \parallel 1> = [ a(\theta)]^{N_3} [\bar{b}(\theta)]^{N^*_3}
\parallel 1>,& \exxi a \cr
&B_{i}\parallel 1> \neq 0,\qquad i = 2, 3 , &\exxi b \cr
&C_{i}\parallel 1> = 0,\qquad i = 2, 3, &\exxi c \cr
&D_{i, j}\parallel 1> \in  \Omega,\qquad i, j= 2, 3, &\exxi d \cr
}$$
}
$N_3$ $(N^*_3)$  being the number of sites in the representation $\{3\}$
$(\{3^*\})$.
In our case  $N_3 =N^*_3= N$.

Following the steps inspired in the NBA, we apply r-times the $B$ operators to
$\parallel 1>$ and build the state
\eqn\exxii{
\Psi (\vec{\mu}) \equiv \Psi (\mu_1, \cdots, \mu_r)=B_{i_1}(\mu_1)\otimes
\ldots \otimes B_{i_r}(\mu_r) X_{i_1,\cdots,i_r} \parallel 1>,
}
$X_{i_1,\cdots,i_r} $ being a $r$-tensor that, together with the values of the
spectral parameters $\mu_1 ,\cdots, \mu_r$, will be determined at the end.

The action  of $A(\mu)$ and $D_{i,i}(\mu)$ on $\Psi$ is found by pushing them
to the right  through the $B_{i_j}(\mu_j)$'s using the commutations rules
\exvii {b,c}. Two types of terms arise  when $A$ and $D_{i,j}$ pass through
$B$'s, the wanted and unwanted terms, similar to the  obtained in the NBA
method. The first one comes from the first terms of \exvii {b,c}. In this type
of terms the $A$ or $D_{i,i}$ and the $B$'s keep their original arguments and
give a state proportional to $\Psi$. The terms coming from the second terms in
\exvii{b,c} are called unwanted since they contain $B_i(\mu)$ and so they never
give a state proportional to $\Psi$; so, they must cancel each other out when
we sum the trace of $T^{alt}$.

The wanted term obtained by application of $A$ is
\eqn\exxiii{
[a(\mu)]^{N_3}
[\bar{b}(\mu)]^{N^*_3}\prod_{j=1}^{r}{g(\mu_j-\mu)}B_{i_1}(\mu_1)\otimes \ldots
\otimes B_{i_r}(\mu_r) X_{i_1,\cdots,i_r} \parallel 1>,
}
and the $k$-th unwanted term
\eqn\exxiv{
\eqalign{
-[a(\mu_k)]^{N_3} [\bar{b}(\mu_k)]^{N^*_3}
\prod_{j=1 \atop j\neq k}^{r}{g({\mu}_{j} - {\mu}_{k})} &
\left( B(u) \tilde{r}^{(2)}({\mu}_{k} - u)  \right)
\otimes {B}({\mu}_{k+1})\otimes \cdots \cr
&\cdots\otimes {B}({\mu}_{r})\otimes {B}({\mu}_{1})   \otimes {B}({\mu}_{k-1})
{M}^{(k-1)} X \parallel 1>, \cr
}}
$M$ being the operator arising by repeated application of \exvii {a}\ ,
\eqn\exxv{
{B}({\mu}_{1})\otimes \cdots \otimes {B}({\mu}_{r})={B}({\mu}_{k+1})\otimes
\cdots
{B}({\mu}_{r})\otimes {B}({\mu}_{1}) \cdots  \otimes {B}({\mu}_{k-1})
{M}^{(k-1)}.
}

The application of the operators $D_{i,j}(\mu)$ to the state $\Psi (\vec{\mu})$
results more complicated. The  wanted term results
\eqn\exxvi{
\eqalign{
\bigl[
&D_{k, j}(\mu)B_{i_1}(\mu_1)\otimes \ldots \otimes B_{i_r}(\mu_r)
X_{i_1,\cdots,i_r} \parallel 1> \bigr]_{wanted}=
\prod_{i=1}^{r}{g(\mu-\mu_i)} \cr
&{R}_{ j_r, a_r}^{(2) a_{r-1}, i_r} (\mu-\mu_r) \cdots
{R}_{  j_2, a_2}^{(2) a_{1}, i_2} (\mu-\mu_2) \cdot
{R}_{ j_1, a_1}^{(2) j, i_1} (\mu-\mu_1)
D_{k, a_r} X_{i_1,\cdots,i_r} \parallel 1>, \cr
}}
where the $R^{(2)}$'s product is taken in the auxiliary space and has the form
\eqn\exxvii{
\Phi(\mu, \vec{\mu})_{a_r, j}\equiv{R}_{ j_r, a_r}^{(2) a_{r-1}, i_r}  \cdots
{R}_{  j_2, a_2}^{(2) a_{1}, i_2}  \cdot
{R}_{ j_1, a_1}^{(2) j, i_1} =
\pmatrix{
\alpha(\mu, \vec{\mu}) & \beta(\mu, \vec{\mu}) \cr
\gamma(\mu, \vec{\mu}) & \delta(\mu, \vec{\mu}) \cr
}.
}

The action of  $D_{k,j}$ with $k\neq j$ on $\parallel 1>$ is not zero. This is
the main difference with the models that can be solved by NBA. Then, we try to
diagonalize the matrix product
\eqn\exxviii{
F(\mu, \vec{\mu})=D(\mu) \cdot \Phi(\mu, \vec{\mu})=
\pmatrix{
A^{(2)}(\mu, \vec{\mu}) & B^{(2)}(\mu, \vec{\mu}) \cr
C^{(2)}(\mu, \vec{\mu}) & D^{(2)}(\mu, \vec{\mu}) \cr
}.
}
By taking the terms in \exxvi\ with $k=j$ and adding them for $k=2$ and $3$, we
obtain the wanted term
\eqn\exxix{
\prod_{j=1}^{r}{g(\mu-\mu_j)}B_{i_1}(\mu_1)\otimes \ldots \otimes
B_{i_r}(\mu_r) \tau_{(2)}(\mu, \vec{\mu})X_{i_1,\cdots,i_r} \parallel 1>,
}
where
\eqn\exxx{
\tau_{(2)}(\mu, \vec{\mu})=\trace(F)=A^{(2)}(\mu, \vec{\mu})+D^{(2)}(\mu,
\vec{\mu}).
}

In the same form, the $k$-th unwanted term results
\eqn\exxxi{
\eqalign{
-\prod_{j=1 \atop j\neq k}^{r}{g({\mu}_{k} - {\mu}_{j})} &
\left( B(\mu) \tilde{r}^{(2)}(\mu-{\mu}_{k} )  \right)
\otimes {B}({\mu}_{k+1})\otimes \cdots \cr
&\cdots\otimes {B}({\mu}_{r})\otimes {B}({\mu}_{1})   \otimes {B}({\mu}_{k-1})
{M}^{(k-1)}   \tau_{(2)}(\mu_k, \vec{\mu}) X \parallel 1>. \cr
}}
The  sum of the wanted terms and the cancelation of the unwanted terms give us
the relations
\eqn\exxxii{
 \tau_{(2)}(\mu, \vec{\mu}) X \parallel 1> =  \Lambda_{(2)}(\mu, \vec{\mu}) X
\parallel 1>
}
and
\eqn\exxxiii{
 \Lambda_{(2)}(\mu_k, \vec{\mu})=
[a(\mu_k)]^{N_3} [\bar{b}(\mu_k)]^{N^*_3}
\prod_{j=1 \atop j\neq k}^{r}{{g({\mu}_{j} - {\mu}_{k}) \over g({\mu}_{k} -
{\mu}_{j})}}.
}
We must now diagonalize \exxxii.

The tensor $X_{i_1,\cdots,i_r},  (i_j =2,3)$ lies in a space with $2^r$
dimensions and  $ \parallel 1> \in \Omega$. Then, the vector $X \parallel 1>$
yields in a space $\Omega^{(2)}$ with $2^{r+N}$ dimensions. In this space, we
take the element
\eqn\exxxiv{
\parallel 1>^{(2)}=\left (\matrix{
1\cr
0\cr
}\right )_1
\otimes
 \cdots
\otimes
\left (\matrix{
1\cr
0\cr
}\right )_r\otimes
|u \bar{s}>_1\otimes\cdots\otimes|u \bar{s}>_N ,
}
which is annihilated by $C^{(2)}(\mu, \vec{\mu})$. (Note that the operators
$\alpha$, $\beta$, $\gamma$ and $\delta$ of  {\exxvii} act on first part of
$\parallel 1>^{(2)}$ and the operators $D_{i,j}$ on the second part ). The
application of the operators $A^{(2)}$ and $D^{(2)}$ gives
\eqna\exxxv{
$$\eqalignno{
&A^{(2)} (\mu, \vec{\mu})\parallel 1>^{(2)}=[b(\mu_k)]^{N_3}
[\bar{b}(\mu_k)]^{N^*_3}\parallel 1>^{(2)}, &\exxxv a \cr
&D^{(2)} (\mu, \vec{\mu})\parallel 1>^{(2)}=\prod_{i=1}^{r}{{1 \over
g(\mu-\mu_i)}}
[a(\mu_k)]^{N_3} [\bar{b}(\mu_k)]^{N^*_3}\parallel 1>^{(2)}, &\exxxv b \cr
}$$}

The important fact is that $F(\mu, \vec{\mu})$ verifies the YBE with the
$R^{(2)}$ matrix given in \exix,
\eqn\exxxvi{
R^{(2)}(\mu-\mu') [F(\mu, \vec{\mu}) \otimes F(\mu', \vec{\mu})]=
 [F(\mu', \vec{\mu}) \otimes F(\mu, \vec{\mu})]R^{(2)}(\mu-\mu'),
}
which, in a second step, permits us to solve the system. From this equation, we
obtain the commutation rules
\eqna\exxxvii{
$$\eqalignno{
&A^{(2)} (\mu) \cdot B^{(2)} (\mu')=g(\mu'-\mu) B^{(2)} (\mu') \cdot A^{(2)}
(\mu )-
h_+(\mu'-\mu) B^{(2)} (\mu ) \cdot A^{(2)} (\mu') ,&\exxxvii a \cr
&D^{(2)} (\mu ) \cdot B^{(2)} (\mu')=g(\mu-\mu') B^{(2)} (\mu') \cdot D^{(2)}
(\mu )-
h_+(\mu-\mu') B^{(2)} (\mu) \cdot D^{(2)} (\mu') .&\exxxvii b \cr
}$$}

In this second step, we build the vector
\eqn\exxxviii{
\Psi^{(2)}(\vec{\lambda},\vec{\mu})=B^{(2)} (\lambda_1, \vec{\mu}) \cdots
B^{(2)} (\lambda_s, \vec{\mu})\parallel 1>^{(2)} .
}
The action of $A^{(2)} (\lambda, \vec{\mu})$ on $\Psi^{(2)}$ gives the wanted
term
\eqn\exxxix{
[b(\lambda)]^{N_3} [\bar{b}(\lambda)]^{N^*_3} \prod_{i=1}^{s}{g(\lambda_i -
\lambda})  B^{(2)}(\lambda_1,\vec{\mu}) \ldots B^{(2)}(\lambda_s,\vec{\mu})
\parallel 1>^{(2)},
}
and the $k$-th unwanted term
\eqn\exxxx{
\eqalign{
-h_{+}(\lambda_k-\lambda ) [b(\lambda_k)]^{N_3} [\bar{b}(\lambda_k)]^{N^*_3} &
\prod_{i=1\atop i\neq k}^{s}
{g(\lambda_i - \lambda_k)}  B^{(2)}(\lambda,\vec{\mu})
B^{(2)}(\lambda_{k+1},\vec{\mu}) \ldots    \cr &\ldots
B^{(2)}(\lambda_{k-1},\vec{\mu}) \parallel 1>^{(2)}. \cr
}}
In the same form, the action of $D^{(2)} (\lambda, \vec{\mu})$ on $\Psi^{(2)}$
gives the wanted term
\eqn\exxxxi{
[b(\lambda)]^{N_3} [\bar{a}(\lambda)]^{N^*_3}
\prod_{i=1}^{s}{g( \lambda}-\lambda_i)
\prod_{j=1}^{r}{1  \over g(\lambda_k-\mu_j)}
B^{(2)}(\lambda_1,\vec{\mu}) \ldots B^{(2)}(\lambda_s,\vec{\mu}) \parallel
1>^{(2)},
}
and the $k$-th unwanted term
\eqn\exxxxii{
\eqalign{
-h_{-}(\lambda-\lambda_k) [b(\lambda_k)]^{N_3} [\bar{a}(\lambda_k)]^{N^*_3} &
\prod_{i=1\atop i\neq k}^{s}{g(\lambda_k - \lambda_i)}
\prod_{j=1}^{r}{1  \over g(\lambda_k-\mu_j)}
B^{(2)}(\lambda,\vec{\mu})
B^{(2)}(\lambda_{k+1},\vec{\mu}) \ldots    \cr &\ldots
B^{(2)}(\lambda_{k-1},\vec{\mu}) \parallel 1>^{(2)}. \cr
}}

The cancelation of the unwanted terms and the sum of the wanted terms, give us
the equations
\eqn\exxxxiii{
\biggl[ { {\bar{a}(\lambda_k)} \over {\bar{b}(\lambda_k)} } \biggr] ^{N^*_3}
\prod_{j=1}^{r}{1  \over g(\lambda_k-\mu_j)}=
\prod_{i=1\atop i\neq k}^{s}{ {g(\lambda_i - \lambda_k)} \over
{g(\lambda_k - \lambda_i)}} \qquad k=1,\ldots,s
}
and
\eqn\exxxxiv{
\Lambda_{(2)}(\mu_k, \vec{\mu})=
\prod_{i=1}^{s}{ g(\lambda_i-\mu_j)}
[b(\mu_k)]^{N_3} [\bar{a}(\mu_k)]^{N^*_3}.
}

Then, by comparing equations \exxxiii\ and \exxxxiv\  and calling
$\bar{g}(\theta)={\bar{a}(\theta) / \bar{b}(\theta)}$, we obtain the coupled
Bethe equations
\eqna\exxxvii{
$$\eqalignno{
&[\bar{g}(\lambda_k)]^{N^*_3} =\prod_{j=1}^{r}{g(\lambda_k-\mu_j)}
\prod_{i=1\atop i\neq k}^{s}{ {g(\lambda_i - \lambda_k)} \over
{g(\lambda_k - \lambda_i)}}, &\exxxvii a \cr
&[g(\mu_k)]^{N_3} =
\prod_{j=1\atop j \neq k}^{r}{ {g(\mu_k -\mu_j)} \over
{g(\mu_j-\mu_k)}}
\prod_{i=1}^{s}{g(\lambda_i-\mu_k)}, &\exxxvii b \cr
}$$}
and the eigenvalue of the trace of  $T^{(alt)}$
\eqn\exxxviii{
\eqalign{
\Lambda(\mu)&=
[a(\mu)]^{N_3} [\bar{b}(\mu)]^{N^*_3}\prod_{j=1}^{r}{g(\mu_j-\mu)}+  \cr
&[b(\mu)]^{N_3}  \prod_{j=1}^{r}{g(\mu-\mu_j)}
\biggl[ [\bar{b}(\mu)]^{N^*_3} \prod_{i=1}^{s}{g(\lambda_i-\mu)} +
[a(\mu)]^{N_3} \prod_{i=1}^{s}{g(\mu-\lambda_i)}
\prod_{j=1}^{r}{{1 \over g(\mu-\mu_j)}} \biggr]  \cr
}}
that is the solution to our problem.

The solution we have obtained, permits us to conjecture the form of the
solution for a non homogeneous chain combining the two elementary
representations of an algebra $A_n$. Each elementary representation introduces
a function $g$ and $\bar{g}$ (that we can call source functions).  Such
solution will have $n$ sets of Bethe equations ( the same number of basic
representations of $A_n$ ). The first and last sets will have in the first
member their respective source functions powered to the number of sites of each
representation and in the second member a product of source functions similar
to those in the right hand side of eqns. \exxxvii{a,b}. The left hand sides of
the rest of the sets of equations are the unity, since there is not source
function. For other representations, we expect a similar scheme with source
functions related to the source functions of the elementary representations
\ref\rrx{Discussions on those points with H. de Vega are acknowledge}. Since
our method has a wide applicability, we will study these points in a
forthcoming paper.

{ \tenbf  Acknowledgements}

We thank to professor H. de Vega for very useful discussions and remarks. A
careful reading of the manuscript by professor J. Sesma is also acknowledged.
This work was partially supported by the Direcci\'{o}n General de
Investigaci\'{o}n Cient\'{\i}fica y T\'{e}cnica, Grant No PB93-0302 and
AEN94-0218

\listrefs
\end{document}